\documentclass[twocolumn,showpacs,preprintnumbers,amsmath,amssymb,floatfix]{revtex4}
\input{epsf}

\usepackage{graphicx}% Include figure files
\usepackage{dcolumn}% Align table columns on decimal point
\usepackage{bm}% bold math

\begin{document}

\title{Production of mesoscopic superpositions with ultracold atoms}
\author{H. T. Ng}
\affiliation{{Clarendon Laboratory, Department of Physics,
University of Oxford, Parks Road, Oxford OX1 3PU, United Kingdom}}
\date{\today}

\begin{abstract}
We study mesoscopic superpositions of two component Bose-Einstein
condensates.  Atomic condensates, with long coherence times, are
good systems in which to study such quantum phenomenon. We show that
the mesoscopic superposition states can be rapidly generated in
which the atoms dispersively interact with the photon field in a
cavity. We also discuss the production of compass states which are
generalized Schr\"{o}dinger cat states. The physical realization of
mesoscopic states is important in studying decoherence and precision
measurement.
\end{abstract}


\maketitle
\section{Introduction}
In 1935, Schr\"{o}dinger proposed a thought-experiment of a
superposition of macroscopically distinct states \cite{Schrodinger}.
Study of macroscopic superpositions is crucial to our understanding
the boundary between the quantum and classical world, and the quest
for physical realization of macroscopic superposition states is
ongoing \cite{Leggett}.

In fact, mesoscopic superpositions of photon field have been
observed in cavity QED experiment by Brune {\it et al.}
\cite{Brune}. They provide a very clear way to create and measure
superposition states of photons in a cavity \cite{Brune}. In this
experiment, a two-level atom is prepared as a superposition of the
excited and ground states. The microwave photon field is a coherent
state $|\alpha\rangle$ of a quantum harmonic oscillator
\cite{Scully}, where the mean number of photons is $|\alpha|^2$. The
two-level atom is sent through the cavity and dispersively interacts
with the microwave field.  In this process, the two opposite phase
shifts of photon field depending on two different states of atom are
acquired. Consequently, a superposition of photon states is
generated.

However, this photon state is a superposition of the two coherent
states with only a few quanta \cite{Brune}. It is still far-reaching
to realize robust mescoscopic superpositions because they are
extremely fragile.  Thus, it is necessary to find a decoherence-free
quantum system or system with a low decoherence rate to generate and
observe such states. Bose-Einstein condensation of atomic gases
\cite{Bradley} is a promising candidate to exhibit this quantum
phenomenon. Many fascinating experiments of atomic Bose-Einstein
condensates (BEC's) have been done in which long coherence times
have been observed \cite{Harber,remark}. This may be used to produce
a superposition of quantum states so a ``big cat'' state would be
experimentally observed.

\begin{figure}[ht]
\includegraphics[height=4.5cm]{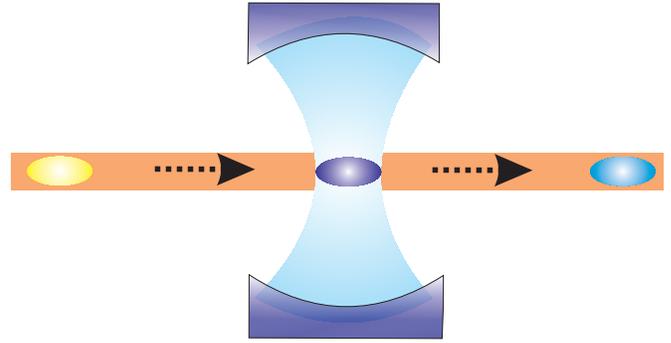}
\caption{ \label{fig1}  Configuration of a cavity and condensate are
shown.  The condensate is transported along a waveguide and moves
toward the cavity.  The atoms dispersively interact with the cavity
field.  The changes of colour of the condensates to denote the
changes of states at different stages. }
\end{figure}

In this paper, we will adopt the idea by Brune {\it et al.}
\cite{Brune} to generate mesoscopic superpositions of BEC's with a
cavity.  Recently, a BEC has been loaded into a magnetic trap using
optical tweezers \cite{Gustavson} and can also be transported via a
magnetic waveguide \cite{Leanhardt}. It paves a way to transfer a
BEC into a cavity.  For a two-component condensate, its ground state
and low lying excitations behave in the same way of a quantum
harmonic oscillator \cite{Ng}. More than that, this ``mesoscopic
quantum harmonic oscillator'' being made by BEC, is more robust and
long-lived than that of the cavity photon field.  Indeed, the
coherence times of the atomic condensates have been measured in the
Ramsey spectroscopy which exceeds 1 s \cite{Harber}. It is therefore
very encouraging that quantum superpositions of this mesoscopic
quantum object will be observed.

We consider a two-component condensate sent through a high-Q
microwave cavity.  The atoms dispersively interact with the photon
field in which no absorption or emission of photons occurs in the
dynamics.  In this paper, we will show that a superposition of the
condensates states with two different relative phases \cite{Hall}
are generated during the dispersive interaction between the atoms
and the photon field. The detection of these superposition states is
also discussed.

We will also investigate the creation of compass states of ultracold
atoms with a cavity. Compass states have been proposed by Zurek
\cite{Zurek} which are ``generalized'' Schrodinger cat states with a
superposition of more than two distinguishable coherent states.
These states can be used to achieve high precision measurement, such
as the detection of small displacements or rotations in phase space
\cite{Toscano}.

\section{System}
\begin{figure}[ht]
\includegraphics[height=5.5cm]{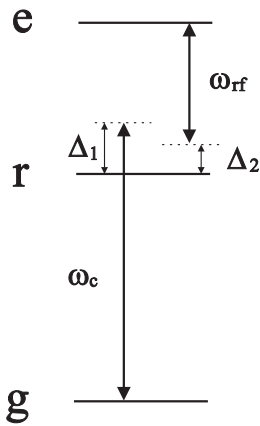}
\caption{ \label{engdiagm1}  Level diagram of the two-photon
transition is shown. The two levels $|r\rangle$ and $|g\rangle$
interacts with a microwave cavity field $\omega_c$ with a detuning
$\Delta_1$. The states $|e\rangle$ and $|r\rangle$ are coupled via a
classical radio-frequency field with a detuning $\Delta_2$. }
\end{figure}

We consider the two hyperfine states
$|e\rangle\equiv|F=2,m_F=1\rangle$  and
$|g\rangle\equiv|F=1,m_F=-2\rangle$  of ${}^{87}$Rb in which the
energy splitting between these two hyperfine levels is about $6.8$
GHz \cite{Matthews}. The states are very robust to magnetic noise
since they have similar magnetic moments \cite{Treutlein1}. The
lower state $|g\rangle$ couples to  an immediate state
$|r\rangle\equiv|F=2,m_F=0\rangle$ via a microwave cavity field with
a detuning $\Delta_1=\omega_{\rm rg}-\omega_c$, for $\omega_{\rm
rg}$ and $\omega_c$ are the energy difference frequencies between
the states $|r\rangle$ and $|g\rangle$, and the microwave frequency
respectively. A classical field with radio frequency $\omega_{\rm
rf}$ is applied to the cavity and no photon is fed into the cavity.
The upper state $|e\rangle$ interacts with the state $|r\rangle$
through the classical radio-frequency (rf) field with a detuning
$\Delta_2=\omega_{\rm er}-\omega_{\rm rf}$, where $\omega_{\rm er}$
is the energy splitting frequency. The two internal states
$|e\rangle$ and $|g\rangle$ are effectively coupled
\cite{Treutlein1} via the two-photon transition \cite{Gentile} as
shown in Fig \ref{engdiagm1}. In this way, the two-photon transition
can be used as a ``quantum switch'' of the interaction of two
internal states and the microwave field through controlling the rf
field.

In the interaction picture, the effective Hamiltonian, describing
the two-photon interaction between the microwave cavity field and
the classical rf field, and the number of $N$ two-level atoms in the
condensates, has the form ($\hbar=1$):
\begin{eqnarray}
H&=&{\Delta}J_{z}-g(a^{\dag}J_{-}+J_{+}a),
\end{eqnarray}
where $a$, $J_z$ and $J_{\pm}$ are the annihilation operator of the
microwave cavity field, atomic inversion and atomic transition
operators for the condensates respectively.  The parameters
$\Delta=\Delta_1+\Delta_2-\Omega^2_{\rm rf}/\Delta_1$ and
$g=\Omega_c\Omega_{\rm rf}/\Delta_1$ are the two-photon detuning and
the effective two-photon Rabi frequency respectively.

The BEC has been confined in a very tight trap with the transverse
size being less than $500$ $\mu$m \cite{Leanhardt}.  The wavelength
of a microwave field is much longer than the size of the
condensates.  Therefore, we have considered that all atoms interact
with the cavity field in the same form.  We also assume that
atom-atom interactions are very small compared to the strength of
atom-light coupling so that we can neglect the nonlinear terms
$\kappa{J^2_z}$ coming from the self-interactions between the atoms
\cite{remark0}.

\subsection{Effective Jaynes-Cummings (JC) Model}
In the low-lying excitations regime, we are able to make the
approximation based on the Holsetin-Primakoff transformation
\cite{Holstein}.  This allows us to approximate the angular momentum
operators into harmonic oscillators: $J_{+}\approx{\sqrt{N}b^\dag}$,
$J_{-}\approx{\sqrt{N}b}$ and $J_z=b^\dag{b}-N/2$ as long as the
excitations are small, i.e., $\langle{b^\dag{b}}\rangle/N{\ll}1$
\cite{Ng}.  This means an atomic coherent state with ``small''
excitations can be approximated to a coherent state of a harmonic
oscillator which can be expressed in the form as \cite{Scully}:
\begin{eqnarray}
|\alpha\rangle&=&e^{-|{\alpha^2}/{2}|}\sum_n\frac{\alpha^n}{(n!)^{1/2}}|n\rangle,
\end{eqnarray}
where $n$ is a quantum number of excitations in the condensates. The
degree of excitation of the atomic coherent state can be adjusted by
a two-photon Rabi pulse \cite{Hall}. The condensates can well be
approximate as a coherent state for which the number of excitations
$|\alpha|^2$ is up to several hundred and the number of atoms
$N\sim{10^4}$ to $10^5$ \cite{Leanhardt}.

In the following discussion, we consider the photon state of the
cavity field is a superposition of the single photon and vacuum
states.  We initially prepare the photon state as:
\begin{equation}
\label{photon-state}
|\psi\rangle=\frac{|0\rangle+|1\rangle}{\sqrt{2}}.
\end{equation}
This state has been experimentally observed in a microwave cavity
\cite{Maitre}. It can be achieved by sending a two-level atom, with
an equally superposition of states $|r\rangle$ and $|g\rangle$,
through a cavity in which the vacuum photon field state $|0\rangle$
is prepared \cite{Maitre}. After a certain time, the photon field
state will become the form in Eq. (\ref{photon-state}).  If we
constraint the photon state in the single excitation and the vacuum
state only such that we are able to represent the bosonic
oscillators in terms of Pauli operators \cite{Sachdev}:
$\sigma^x=a^\dag+a$, $\sigma^y=i(a^\dag-a)$ and
$\sigma^z=1-2a^\dag{a}$.

After the above transformations, we can readily obtain an effective
Hamiltonian as:
\begin{eqnarray}
H_{\rm JC}&=&{\Delta}b^\dag{b}+g\sqrt{N}(b^\dag\sigma_-+\sigma_+b).
\end{eqnarray}
We can see that this Hamiltonian has the same form of the well-known
Jaynes-Cumming (JC) model \cite{Scully}.  This effective Hamiltonian
is valid as long as no further excitations of the photon field is
made in the dynamics.  Thus, it is very good to describe the
dispersive interaction between atoms and microwave field in which no
net photon is either absorbed or created during the process.

It is noteworthy that the Rabi-coupling strength is vastly enhanced
by a factor of $\sqrt{N}$.  This effective JC Rabi-coupling strength
$g\sqrt{N}$ can be increased with 100 times for $N$ ${\approx}$
$10^4$. We estimate this effective Rabi frequency can be high as 100
kHz if $g\approx{1}$ kHz.  The strong atom-photon coupling strength
is crucial to increase the efficiency of the cat states production.

This JC model can be exactly diagonalized which allows us to solve
the problem analytically. If the detuning $\Delta$ is large enough
with $4g^2Nn/\Delta^2{\ll}1$, the eigenvectors can be expressed in
this approximate form as \cite{Scully}:
$(i)~|+\rangle\approx|1,n-1\rangle$ and
$(ii)~|-\rangle\approx|0,n\rangle$, and the corresponding
eigenvalues are $(i)~E_+\approx(n-1){\Delta}-{g^2Nn}/{\Delta}$ and
$(ii)~E_-\approx {n\Delta}+{g^2Nn}/{\Delta}$.

\section{Generation of Schr\"{o}dinger Cat State}
Now we propose a scheme to generate the cat state with only a few
steps, where the atoms are dispersively coupled to microwave field.
We consider the initial state as a product state of the photon and
the atoms, i.e.,
\begin{equation}
|\Psi(0)\rangle=\Bigg(\frac{|0\rangle+|1\rangle}{\sqrt{2}}\Bigg){\otimes}|\alpha\rangle.
\end{equation}
Then, the condensate is transported through the cavity, and the
interaction between the atoms and microwave field is switched on for
the duration time $t^*$. The state will become \cite{Brune}
\begin{equation}
\label{cat}
|\Psi(t^*)\rangle=\frac{1}{\sqrt{2}}(|1,\tilde{\alpha}e^{i\phi}\rangle+|0,\tilde{\alpha}e^{-i\phi}\rangle),
\end{equation}
for $\tilde{\alpha}={\alpha}e^{-i{\Delta}t^*}$ and
$\phi=g^2Nt^*/\Delta$.  The smooth transport of atoms through a
magnetic waveguide has also been demonstrated which shows negligible
atoms losses during the transfer \cite{Leanhardt}.

We can make the unitary transformation of the photon state
$U_{\pi/2}$ with a two-level atom as follows \cite{Maitre}:
$|1\rangle\rightarrow(|1\rangle+|0\rangle)/\sqrt{2}$ and
$|0\rangle\rightarrow(|0\rangle-|1\rangle)/\sqrt{2}$.  This can be
achieved by interacting the photon field with a two-level atom
appropriately \cite{Maitre}.  The state is then written as
%\begin{widetext}
\begin{eqnarray}
|\Psi\rangle&=&|1\rangle|\tilde{\Psi}_1\rangle+|0\rangle|\tilde{\Psi}_2\rangle,
\end{eqnarray}
where
\begin{equation}
|\tilde{\Psi}_{1,2}\rangle=\Bigg(\frac{|\tilde{\alpha}e^{i\phi}\rangle{\mp}|\tilde{\alpha}e^{-i\phi}\rangle}{\sqrt{2}}\Bigg).
\end{equation}
%\end{widetext}
A Schr\"{o}dinger cat state of condensates is readily obtained by
either measuring the photon state $|1\rangle$ or $|0\rangle$. The
two possible superposition states are
\begin{equation}
\label{cat1}
|\tilde{\Psi}_{1,2}\rangle=\frac{|\tilde{\alpha}e^{i\phi}\rangle{\mp}|\tilde{\alpha}e^{-i\phi}\rangle}{{\tilde{N}_{1,2}}},
\end{equation}
where
$\tilde{N}^2_{1,2}=2\mp{2}e^{-|\tilde{\alpha}|^2(1-\cos2\phi)}\cos(|\tilde{\alpha}^2|\sin2\phi)$.
Physically speaking, this cat state is a superposition of the states
with two different relative phases of the two-component condensates.
The lifetime $T$ of the cat state is roughly equal to $2T_r/D^2$ as
$D\gg1$ \cite{Brune}, where $T_r$ is the characteristic damping time
of the condensates and $D=2|\tilde{\alpha}|$ is ``distance'' between
two coherent states \cite{Brune}.  The characteristic damping time
of the BEC is about the coherence lifetime of the condensates $T_c$.
Obviously, the cat state lifetime is significantly shortened if the
magnitude of $|\tilde{\alpha}|^2$ is very large. But the coherence
time of the atomic condensates ($T_c$ ${\sim}$ 1 to 2 s) is
relatively long \cite{remark}, we are able to estimate the cat
state's lifetime $T$ is about 1 to 10 ms if the number of
excitations is up to several hundred.

Next we discuss how to detect such superposition states in
condensates. We note that the photon and the condensates are
entangled in Eq. (\ref{cat}), therefore the photon state can be used
to probe the state of atomic condensates. This enables us to measure
the superposition states through detecting the photon state. We
consider the cat state in Eq. (\ref{cat}) with $\phi=\pi/2$.  Let
the system undergo a free evolution $(H_0=\Delta'b^\dag{b})$ for a
small amount of time $\delta{t}$, where
$\Delta'=\Delta_1+\omega_{er}$. We switch on the JC interaction
again for a duration $t'=\pi\Delta/2g^2N$ to undo the cat state.
After performing $U_{\pi/2}$ on the photon state, up to a global
phase, the quantum state will become \cite{Toscano}:
\begin{equation}
|\Psi_f\rangle=\frac{1}{2}[(e^{2i\Delta'|\tilde{\alpha}|^2\delta{t}}-1)|1,\tilde{\alpha}'\rangle+(e^{2i\Delta'|\tilde{\alpha}|^2\delta{t}}+1)|0,\tilde{\alpha}'\rangle].
\end{equation}
for $\tilde{\alpha}'=\tilde{\alpha}e^{-i\Delta{t'}}$. The frequency
is amplified by a factor $|\tilde{\alpha}|^2$ as the probability of
single photon state is
$P_1=[1-\cos(2|\tilde{\alpha}'|^2\Delta'\delta{t})]/2$
\cite{Toscano}. The probability of photon state can be measured by
means of a two-level atom \cite{Maitre}.  Indeed, this method has
been proposed to perform Heisenberg limited precision measurement
\cite{Toscano} to detect the small rotation of the state in the
phase space.

\section{Compass state Production}
Having discussed how to generate Schr\"{o}dinger cat states, we
proceed to study another kind of superposition states which is
called the compass state \cite{Zurek,Toscano}.  It is a
generalization to the usual Schr\"{o}dinger cat state which is a
superposition of two coherent states.  These compass states play an
important role in the study of sub-Planck structure \cite{Zurek}.

We follow the similar method in generating cat states to make a
compass state.  We first create the state in Eq. (\ref{cat1}) with
$\phi=\pi/2$, and prepare the photon state in the cavity in the form
of Eq. (\ref{photon-state}).  Then,  we switch on the effective JC
interaction with a duration $t''=\pi\Delta/4g^2N$. The four phase
shifts of the coherent state are yielded as
$|\tilde{\alpha}^*e^{{\pm}i\pi/4}\rangle$ and
$|\tilde{\alpha}^*e^{{\pm}i3\pi/4}\rangle$, for
$\tilde{\alpha}^*=\tilde{\alpha}e^{-i\Delta{t''}}$.  The unitary
transformation of the photon state $U_{\pi/2}$ is then performed.
After measuring the photon state $|0\rangle$ or $|1\rangle$ of the
cavity, the state finally becomes
\begin{widetext}
\begin{equation}
|\tilde{\Psi}_{1,2}'\rangle\approx\frac{|{\tilde{\alpha}^*}e^{i\pi/4}\rangle\mp|{\tilde{\alpha}^*}e^{-i\pi/4}\rangle+|{\tilde{\alpha}^*}e^{3i\pi/4}\rangle\mp|
{\tilde{\alpha}^*}e^{-3i\pi/4}\rangle}{2}.
\end{equation}
\end{widetext}
The states $|\tilde{\Psi}_1'\rangle$ and $|\tilde{\Psi}_2'\rangle$
corresponds to the states after the measurement of two photon states
$|0\rangle$ and $|1\rangle$ respectively. The state
$|\tilde{\Psi}_{1,2}'\rangle$ is a superposition of four different
coherent states.  The method here can be generalized to produce a
superposition of more than four distinguishable coherent states.
The successful realization of compass states would allow us to
deepen the understanding of the decoherence problems \cite{Zurek}.

\section{Discussion}
In this paper, we have proposed a method to generate and observe a
long-lived mesoscopic superpositions of two-component BEC's with a
high-Q microwave cavity.  The cat states can be quickly generated in
around 100 $\mu$s if $g$ can attain $1$ kHz or higher.  We require
the photon lifetime is much longer than the time scale of the cat
states production.

In addition, we point out that our scheme to generate cat states
with a cavity is much more efficient than other schemes using
Bose-Einstein condensates \cite{Cirac}.   In those schemes, using
two-component condensates, the generated time is proportional to the
self-interaction strength which is very weak \cite{remark0}. This
requires a much longer production time and thus it is very
unfavorable to generate cat states.

H.T.N. is grateful to Keith Burnett, Jacob Dunningham and David
Hallwood for their useful comments and discussion. H.T.N. thanks the
Croucher Foundation for support.


\begin{thebibliography}{99}
\bibitem{Schrodinger}
E. Schr\"{o}dinger, Naturwissenschaften {\bf 23}, 807 (1935).

\bibitem{Leggett}
A. J. Leggett, J. Phys.:Condens. Matter {\bf 14}, R415-R451 (2002).

\bibitem{Brune}
M. Brune {\it et al.}, Phys. Rev. Lett. {\bf 77}, 4887 (1996).

\bibitem{Scully}
M. O. Scully and M. S. Zubairy, {\it Quantum Optics} (Cambridge
University Press, Cambridge, 1997).


\bibitem{Bradley}
C. C. Bradley {\it et al.}, Phys. Rev. Lett. {\bf 75}, 1687 (1995);
K. B. Davis {\it et al.}, {\it ibid.} {\bf 75}, 3969 (1995); M. H.
Anderson {\it et al.}, Science {\bf 269}, 198 (1995).

\bibitem{Harber}
D. M. Harber {\it et al.}, Phys. Rev. A {\bf 66}, 053616 (2002).

\bibitem{remark}
The coherence of the two component condensates, with hyperfine
states $|F=1,{m_F}=-1\rangle$ and $|F=2,m_F=1\rangle$, is measured
using the Ramsey spectroscopy \cite{Harber}.  The population of
atoms in $|F=2,m_F=1\rangle$ against time $t$ is measured.  The
coherence time $T_c$ is defined as the population decaying with time
$t$ as $e^{-(t/T_c)^2}$.


\bibitem{Gustavson}
T. L. Gustavson {\it et al.}, Phys. Rev. Lett {\bf 88}, 020401
(2002).



\bibitem{Leanhardt}
A. E. Leanhardt {\it et al.}, Phys. Rev. Lett. {\bf 89}, 040401
(2002).




\bibitem{Ng}
H. T. Ng and P. T. Leung, Phys. Rev. A {\bf 71}, 013601 (2005); H.
T. Ng and K. Burnett, {\it ibid.} {\bf 75}, 023601 (2007).


\bibitem{Hall}
D. S. Hall {\it et al.}, Phys. Rev. Lett. {\bf 81},  1543 (1998).



\bibitem{Zurek}
W. H. Zurek, Nature {\bf 412}, 712 (2001).

\bibitem{Toscano}
F. Toscano {\it et al.}, Phys. Rev. A {\bf 73}, 023803 (2006).

\bibitem{Matthews}
M. R. Matthews {\it et al.}, Phys. Rev. Lett. {\bf 81}, 243 (1998).

\bibitem{Treutlein1}
P. Treutlein {\it et al.}, quant-ph/0605163.

\bibitem{Gentile}
T. R. Gentile, B. J. Hughey and D. Kleppner, Phys. Rev. A {\bf 40},
5103 (1989).



\bibitem{remark0}
The interaction strength $\kappa$ is equal to
$\kappa_{e}+\kappa_{g}-2\kappa_{eg}$, for $\kappa_{e(g)}$ and
$\kappa_{eg}$ are the self-interaction strengths of the state $e(g)$
and inter-component strength.  Since the interaction strengths
$\kappa_{e(g)}$ and $\kappa_{eg}$ are very close, the resulting
self-interaction between the two component condensates are very
weak.


\bibitem{Holstein}
T. Holstein and H. Primakoff, Phys. Rev., {\bf 58}, 1098 (1949).





\bibitem{Maitre}
X. Ma\^{i}tre {\it et al.}, Phys. Rev. Lett. {\bf 79}, 769 (1997).

\bibitem{Sachdev}
S. Sachdev, {\it Quantum Phase Transitions} (Cambridge University
Press, Cambridge, 1999).



\bibitem{Cirac}
J. I. Cirac {\it et al.}, Phys. Rev. A {\bf 57}, 1208 (1998); D.
Gordon and C. M. Savage, {\it ibid.} {\bf 59}, 4623 (1999); J.A.
Dunningham and K. Burnett, J. Mod. Opt. {\bf 48}, 1837 (2001).







\end{thebibliography}
\end{document}